\documentclass[preprint,10pt]{elsarticle}
\usepackage{amssymb}
\begin{document}
\begin{frontmatter}
\title{Generalized Gibbons-Hawking-York term for $f(R)$ gravity}
\author{Ahmed Alhamzawi$^*$}
\author{Rahim Alhamzawi$^{\dag}$}
\address{
$^*$Department of Mathematics,
College of Computer Science and Mathematics,
Al-Qadisiyah University,
Iraq\\
$^{\dag}$ Statistics Department, Al-Qadisiyah University, Iraq
\\$^*$ahmzawi@gmail.com
}

\begin{abstract}
A generalization to the Gibbons-Hawking-York boundary term for metric $f(R)$ gravity theories is introduced. A redefinition of the Gibbons-Hawking-York term is proposed. The proposed new definition is used to derive a consistent set of field equations and is extended to metric $f(R)$ gravity theories. The surface terms in the action are gathered into a total variation of some quantity. A total divergence term is added to the action to cancel these terms. Finally, the new definition is proven to demand no restrictions on the value of ${\delta g}_{ab}$ or ${\partial}_{c}{\delta g}_{ab}$ on the boundary.
\end{abstract}
\end{frontmatter}

\section{Introduction}
There are three popular formulations for $f(R)$ gravity theories: metric $f(R)$ gravity, metric-affine $f(R)$ gravity and Palatini $f(R)$ gravity. Metric $f(R)$ has been the most popular of the three, mainly because it passes all the observational and theoretical constrains \cite{faraoni2008f}. Recently, metric $f(R)$ gravity has proven quite useful in providing a toy model for studying the problem of dark energy. One problem with metric $f(R)$ that has received less attention is the problem of fixing the surface terms on the boundary when deriving the field equations. One possible reason for this is that unlike in General Relativity, the boundary terms do not consist of a total variation of some quantity. Therefore, it is not possible to add a total divergence to the action in order to heal the problem and arrive at a well meaning variational principle \cite{sotiriou2010f}. In this paper, we will show that in fact the boundary terms can be expressed as a total variation of some quantity. We will present a redefinition of the Gibbons-Hawking-York \cite{gibbons1977action,york1972role} boundary term to cancel these surface terms and arrive at a consistent set of field equations. It has been widely accepted that the modification of the gravitational action with respect to the surface term by adding the Gibbons-Hawking-York term to the total action is desirable because it cancels the terms involving $\partial{\delta g}_{ab}$, and so setting ${\delta g}_{ab}=0$ becomes sufficient to make the action stationary. In order to see the role Gibbons-Hawking-York boundary term we must first review its role in general relativity. If we consider the Einstein-Hilbert action \cite{wald2010general}
    \begin{eqnarray}\label{Einstein-Hilbert action}
        S_{EH}=\frac{1}{2\kappa}\int{d^{4}x}\sqrt{-g}R,
    \end{eqnarray}
where $R=R_{ab}g^{ab}$ is the Ricci scalar.

The variation of (\ref{Einstein-Hilbert action}) with respect to $\delta{g}_{ab}$ gives
    \begin{eqnarray}\label{Einstein-Hilbert action variation 1}
        \delta S_{EH}=\frac{1}{2\kappa}\int{d^{4}x}(\delta\sqrt{-g}R+\sqrt{-g}\delta R),
    \end{eqnarray}

using $\delta\sqrt{-g}=-\frac{1}{2}\sqrt{-g}{g}_{ab}\delta g^{ab}$ and $\delta{R}_{ab}={\nabla}_{c}{\delta\Gamma}^{c}_{ab}-{\nabla}_{b}{\delta\Gamma}^{c}_{ac}$, the variation $\delta S_{EH}$ becomes
    \begin{eqnarray}\label{Einstein-Hilbert action variation 2}
        \delta S_{EH}=
        &&
        \int_{\nu}{d^{4}x}
        (
        -\frac{1}{2}R g_{ab}\sqrt{-g}\delta{g}^{ab}
        +R_{ab}\sqrt{-g}{\delta g}^{ab}
        \nonumber\\&&
        +\sqrt{-g}g^{ab}({\nabla}_{c}{\delta\Gamma}^{c}_{ab}-
        {\nabla}_{b}{\delta\Gamma}^{c}_{ac})
        ).
    \end{eqnarray}
Finally, using the fact that the covariant derivative of the metric tensor vanishes in general relativity, we can rewrite the third term as the divergence if a vector field $J^{c}$, where $J^{c}=g^{ab}{\delta\Gamma}^{c}_{ab}-g^{ac}{\delta\Gamma}^{b}_{ab}$. Using Gauss-Stokes theorem \cite{poisson2004relativist}, the variation (\ref{Einstein-Hilbert action variation 2}) becomes
    \begin{eqnarray}\label{Einstein-Hilbert action variation 3}
        \delta S_{EH}=
        &&
        \int_{\nu}{d^{4}x}\sqrt{-g}
        (
        R_{ab}-\frac{1}{2}R g_{ab}
        )
        \delta {g}^{ab}
        \nonumber\\&&
        +
        \oint_{\partial\mathcal\nu}{d^{3}x}\sqrt{|h|}
        n_{c}J^{c}
        ,
    \end{eqnarray}
where $h$ is the induced metric tensor associated with hypersurface and $n_{c}$ is the normal unit vector on hypersurface $\partial\nu$. The first term in (\ref{Einstein-Hilbert action variation 3}) is the familiar Einstein field equations multiplied by $\delta{g}^{ab}$, however in order to derive the field equations, the action must be stationary and consequently the second term must vanish. We can cancel the second term by adding a boundary term to the total action in (\ref{Einstein-Hilbert action}) that cancels this surface term.
\newline\newline
Using the definition of the christoffel symbol (i.e. ${\Gamma}^{a}_{bc}=\frac{1}{2}g^{ad}({\partial}_{b}{g}_{cd}+{\partial}_{c}{g}_{bd}-{\partial}_{d}{g}_{bc})$), $J_{c}$ becomes
    \begin{eqnarray}\label{Divergence Term 1}
        J_{c}=g^{ab}({\nabla}_{b}{\delta g}_{ca}-{\nabla}_{c}{\delta g}_{ba}),
    \end{eqnarray}

substituting the completeness relation $g^{ab}=h^{ab}-n^{a}n^{b}$, so that
    \begin{eqnarray}\label{Divergence Term 2}
        n^{c}J_{c}
        &&
        =
        n^{c}(h^{ab}-n^{a}n^{b})({\nabla}_{b}{\delta g}_{ca}-{\nabla}_{c}{\delta g}_{ba})
        \nonumber\\&&
        =
        n^{c}h^{ab}{\nabla}_{b}{\delta g}_{ca}-n^{c}h^{ab}{\nabla}_{c}{\delta g}_{ba}
        \nonumber\\&&
        -n^{a}n^{b}n^{c}{\nabla}_{b}{\delta g}_{ca}+n^{a}n^{b}n^{c}{\nabla}_{c}{\delta g}_{ba}.
    \end{eqnarray}
    The third and fourth terms vanish because of the antisymmetry of $({\nabla}_{b}{\delta g}_{ca}-{\nabla}_{c}{\delta g}_{ba})$, Consequently we obtain
    \begin{eqnarray}\label{Divergence Term 3}
        \left.n^{c}J_{c}\right|_{\partial\nu}=n^{c}h^{ab}({\partial}_{b}{\delta g}_{ca}-{\partial}_{c}{\delta g}_{ba}).
    \end{eqnarray}

    Since $\delta{g}_{ab}$ vanishes everywhere on $\partial\nu$, its tangential derivative must also vanishes. It follows that $h^{ab}{\partial}_{b}{\delta g}_{ca}=0$ and we obtain
     \begin{eqnarray}\label{Divergence Term 4}
        \left.n^{c}J_{c}\right|_{\partial\nu}=-n^{c}h^{ab}{\partial}_{c}{\delta g}_{ba},
    \end{eqnarray}
substituting (\ref{Divergence Term 4}) in (\ref{Einstein-Hilbert action variation 3}), the last term becomes
    \begin{eqnarray}\label{Einstein-Hilbert action variation 4}
        -\oint_{\partial\mathcal\nu}{d^{3}x}\sqrt{|h|}
        n^{c}h^{ab}{\partial}_{c}{\delta g}_{ba}
        .
    \end{eqnarray}
Now, the trick is to construct a term that cancels this term. Gibbons, Hawking and York proposed adding the trace of the extrinsic curvature of the boundary to the action. If we consider the Gibbons-Hawking-York boundary term
    \begin{eqnarray}\label{Gibbons-Hawking-York action}
        S_{GHY}=2\oint_{\partial\mathcal\nu}{d^{3}x}\sqrt{|h|}
        K
        ,
    \end{eqnarray}
where $K={\nabla}_{c}n^{c}$ is intrinsic curvature, using $\left.{\delta K}\right|_{\partial\nu}=n^{c}h^{ab}{\partial}_{c}{\delta g}_{ba}$, the variation of this term gives
    \begin{eqnarray}\label{Gibbons-Hawking-York variation 1}
        {\delta S}_{GHY}=\oint_{\partial\mathcal\nu}{d^{3}x}\sqrt{|h|}
        n^{c}h^{ab}{\partial}_{c}{\delta g}_{ba}
        ,
    \end{eqnarray}

Notice that this boundary term cancels (\ref{Einstein-Hilbert action variation 4}) totally. The role of the Gibbons-Hawking-York is crucial when considering the Einstein-Hilbert action since it is the basis for the most elementary variational principle from which the field equations of general relativity can be defined. However, the use of the Einstein-Hilbert action is appropriate only when the underlying spacetime manifold $\mathcal{\nu}$ is closed, i.e. a manifold which is both compact and without boundary. In the event that the manifold has a boundary $\partial\mathcal{\nu}$, the action should be supplemented by a boundary term so that the variational principle is well-defined.

\section{Redefining the function of Gibbons-Hawking-York term}

Generally, the Gibbons-Hawking-York term is defined as the term added to the total action to cancel the surface terms if the action is to be stationary. However, if we redefine the Gibbons-Hawking-York term as the term add to the Ricci scalar, such that

    \begin{eqnarray}\label{Total Action}
        &&
        \int{d^{4}x}\sqrt{-g}R
        +2\oint_{\partial\mathcal\nu}{d^{3}x}\sqrt{|h|}K
        \nonumber\\&&
        =
        \int{d^{4}x}\sqrt{-g}\tilde{R}
        ,
    \end{eqnarray}
where $\tilde{R}=R-\phi$ and $\phi=2{\nabla}_{c}(h^{ab}\Gamma^{c}_{ab})$. The variation of (\ref{Total Action}) gives
    \begin{eqnarray}\label{Einstein-Hilbert action variation 9}
        \delta\int{d^{4}x}\sqrt{-g}\tilde{R}
        &&
        =
        \int{d^{4}x}\sqrt{-g}
        (
        \delta{R}
        -
        \delta\phi
        -\frac{1}{2}g_{ab}{\delta g}^{ab}\tilde{R}
        )
        \nonumber\\&&
        =
        \int{d^{4}x}\sqrt{-g}
        (
        R_{ab}-\frac{1}{2}g_{ab}R
        ){\delta g}^{ab}
        \nonumber\\&&
        +
        \oint{d^{3}x}\sqrt{|h|}
        (
        J^{c}n_{c}-2h^{ab}\delta\Gamma^{c}_{ab}n_{c}
        )
        ,
    \end{eqnarray}
where we have used
    \begin{eqnarray}\label{Einstein-Hilbert action variation 10}
        &&
        \int{d^{4}x}\sqrt{-g}\delta\phi-\int{d^{4}x}\sqrt{-g}\frac{1}{2}g_{ab}{\delta g}_{ab}\phi
        \\&&
        =
        \delta\int{d^{4}x}(2\sqrt{-g}{\nabla}_{c}(h^{ab}\Gamma^{c}_{ab}))
        =
        2\oint{d^{3}x}\sqrt{|h|}
        h^{ab}\delta\Gamma^{c}_{ab}n_{c}.
        \nonumber
    \end{eqnarray}
Using the fact that the quantity $J^{c}n_{c}-h^{ab}\delta\Gamma^{c}_{ab}n_{c}$ vanishes at the boundary, the variation of (\ref{Total Action}) gives the familiar field equations without boundary terms, i.e.
    \begin{eqnarray}\label{Einstein-Hilbert action variation 10}
        \delta\int{d^{4}x}\sqrt{-g}\tilde{R}=
        \int{d^{4}x}\sqrt{-g}
        (
        R_{ab}-\frac{1}{2}R g_{ab}
        )
        {\delta g}^{ab}
        ,
    \end{eqnarray}
It's clear that the modification considered in (\ref{Total Action}) does the same job as the Gibbons-Hawking-York term without the need to consider the vanishing of ${\delta g}_{ab}$ at the boundary of the hypersurface. Although this may seem simple, in the next section we show that this new definition is crucial in deriving the field equations for f(R) metric gravity without the need to worry about any surface terms that need be to canceled.
\section{Variational principle in metric $f(R)$ gravity}

If we consider the modified $f(R)$ gravity action
    \begin{eqnarray}\label{f(R) action}
        S_{R}=\int{d^{4}x}\sqrt{-g}f(R),
    \end{eqnarray}
the variation of this action
    \begin{eqnarray}\label{Einstein-Hilbert action variation 5}
        \delta S_{R}=\int{d^{4}x}(\delta\sqrt{-g}f(R)+\sqrt{-g}f^{\prime}(R){\delta R}),
    \end{eqnarray}
using ${\delta R}={\delta g}^{ab}R_{ab}+g_{ab}{\nabla}_{c}{\nabla}^{c}({\delta g}^{ab})-{\nabla}_{a}{\nabla}_{b}({\delta g}^{ab})$ and
$\delta\sqrt{-g}=-\frac{1}{2}\sqrt{-g}{g}_{ab}\delta g^{ab}$, (\ref{Einstein-Hilbert action variation 5}) becomes
    \begin{eqnarray}\label{Einstein-Hilbert action variation 6}
        \delta S_{R}
        &&
        =
        \int{d^{4}x}\sqrt{-g}
        (
        f^{\prime}(R){\delta g}^{ab}R_{ab}+f^{\prime}(R)g_{ab}{\nabla}_{c}{\nabla}^{c}({\delta g}^{ab})
        \nonumber\\&&
        -f^{\prime}(R){\nabla}_{a}{\nabla}_{b}({\delta g}^{ab})-\frac{1}{2}f(R)g_{ab}{\delta g}^{ab}
        ).
    \end{eqnarray}
    To derive the field equations we must reexpress the second and third terms as ${\delta g}^{ab}$ multiplied by some quantity. To do that we define
    \begin{eqnarray}\label{W}
        {W}^{c}
        &&
        =
        {f}^{\prime}(R)[{g}_{ab}{g}^{cd}{\nabla}_{d}{\delta g}^{ab}-{\nabla}_{a}{\delta g}^{ca}]
        \nonumber\\&&
        -{\delta g}^{ab}[{g}_{ab}{g}^{cd}{\nabla}_{d}{f}^{\prime}(R)-{\delta}^{c}_{b}{\nabla}_{a}{f}^{\prime}(R)]
        .
    \end{eqnarray}
Rewriting the second and third terms in (\ref{Einstein-Hilbert action variation 6}) in terms of $\nabla_{c}W^{c}$ and using Gauss-Stokes theorem on the last term, (\ref{Einstein-Hilbert action variation 6}) becomes
    \begin{eqnarray}\label{Einstein-Hilbert action variation 7}
        \delta S_{R}
        &&
        =
        \int{d^{4}x}\sqrt{-g}
        (
        f^{\prime}(R)R_{ab}+g_{ab}{\nabla}_{c}{\nabla}^{c}(f^{\prime}(R))
        \nonumber\\&&
        -{\nabla}_{a}{\nabla}_{b}(f^{\prime}(R))-\frac{1}{2}f(R)g_{ab}
        ){\delta g}^{ab}
        \nonumber\\&&
        +
        \oint_{\partial\mathcal\nu}{d^{4}x}\sqrt{|h|}{n}_{c}W^{c}
        .
    \end{eqnarray}
Evaluating the quantity $W^{c}$ at the boundary yields
    \begin{eqnarray}\label{Divergence 8}
        \left.W^{c}\right|_{\partial\nu}=-f^{\prime}(R)n^{c}h^{ab}{\partial}_{c}{\delta g}_{ba}
        .
    \end{eqnarray}
If the action is to be stationary, $W^{c}$ must be canceled with boundary term. One way to do that is to add the Gibbons-Hawking-York term multiplied by $f^{\prime}(R)$ to action in (\ref{f(R) action}), such that

    \begin{eqnarray}\label{Gibbons-Hawking-York action 2}
        S_{R}+S_{GHY mod}
        &&
        =
        \int{d^{4}x}\sqrt{-g}f(R)
        \nonumber\\&&
        +
        2\oint_{\partial\mathcal\nu}{d^{3}x}\sqrt{|h|}
        f^{\prime}(R)K
        .
    \end{eqnarray}

However, the variation of this term requires setting $\delta R=0$ on the boundary of the hypersurface \cite{guarnizo2010boundary}, consequently this creates a strong condition on how ${\delta g}^{ab}$ is varied near the hypersurface since $\partial_{c}{\delta g}^{ab}$ is no longer arbitrary on the boundary. To overcome this restriction we use our new definition of the Gibbons-Hawking-York and replace $f(R)$ by $f(\tilde{R})=f(R-\phi)=\sum^{\infty}_{n=0}(-\phi)^{n}\frac{d^{n}}{dR^{n}}f(R)$, where $\phi$ is now defined such that $\delta\phi={\nabla}_{c}\gamma^{c}_{ab}{\delta g}^{ab}-{\delta g}^{ab}{\nabla}_{c}\gamma^{c}_{ab}$, with the operator $\gamma^{c}_{ab}={g}_{ab}{\nabla}^{c}-{\delta}_{a}^{c}{\nabla}_{b}$. The variation of (\ref{f(R) action}) gives
    \begin{eqnarray}\label{f(R) action 2}
        \delta S_{R}
        &&
        =
        \delta\int{d^{4}x}\sqrt{-g}\sum^{\infty}_{n=0}(-\phi)^{n}\frac{d^{n}}{dR^{n}}f(R)
        \nonumber\\&&
        =
        \delta\int{d^{4}x}\sqrt{-g}({\delta R}-{\delta\phi})f^{\prime}(\tilde{R})
        \nonumber\\&&
        =
        \delta\int{d^{4}x}\sqrt{-g}
        (
        (
        {\delta g}^{ab}{}R_{ab}
        +
        {g}^{ab}{\delta R}_{ab}
        -
        {\nabla}_{c}
        \gamma^{c}_{ab}{\delta g}^{ab}
        \nonumber\\&&
        +
        {\delta g}^{ab}{\nabla}_{c}\gamma^{c}_{ab}
        )
        f^{\prime}(\tilde{R})
        -
        \frac{1}{2}f(\tilde{R})g_{ab}{\delta g}^{ab}
        )
        .
    \end{eqnarray}

Using the definition of $\gamma^{c}_{ab}$, this variation becomes
    \begin{eqnarray}\label{Einstein-Hilbert action variation 9}
        \delta S_{R}
        &&
        =
        \int{d^{4}x}\sqrt{-g}
        (
        f^{\prime}(\tilde{R})R_{ab}+g_{ab}{\nabla}_{c}{\nabla}^{c}(f^{\prime}(\tilde{R}))
        \nonumber\\&&
        -{\nabla}_{a}{\nabla}_{b}(f^{\prime}(\tilde{R}))-\frac{1}{2}f(\tilde{R})g_{ab}
        ){\delta g}^{ab}
        ,
    \end{eqnarray}

for an arbitrary variation ${\delta g}^{ab}$, we get the field equation

    \begin{eqnarray}\label{f(R) field equations 2}
        &&
        f^{\prime}(\tilde{R})R_{ab}+g_{ab}{\nabla}_{c}{\nabla}^{c}(f^{\prime}(\tilde{R}))
        \nonumber\\&&
        -{\nabla}_{a}{\nabla}_{b}(f^{\prime}(\tilde{R}))-\frac{1}{2}f(\tilde{R})g_{ab}
        =0
    \end{eqnarray}

\section{Conclusions}
We have redefined the Gibbons-Hawking-York term as the term add to the Ricci scalar and have shown that this definition works just as well as the original Gibbons-Hawking-York in illuminating any possible boundary terms so that we can have a well defined stationary action. This derivation requires no restrictions on the value of ${\delta g}_{ab}$ or $\partial{\delta g}_{ab}$ on the boundary of the manifold. Although, unlike the familiar metric $f(R)$ field equation \cite{sotiriou2010f}, (\ref{f(R) field equations 2}) is a differential equation for $f(R-\phi)$. Rewriting the variation (\ref{f(R) action 2}) in terms of $f(R)$ yields
    \begin{eqnarray}\label{f(R) action variation 2}
        \delta S_{R}
        &&
        =
        \int{d^{4}x}\sqrt{-g}
        (
        {\delta R}f^{\prime}(R)
        -\frac{1}{2}f(R)g_{ab}{\delta g}^{ab}
        \nonumber\\&&
        -{\delta \phi}f^{\prime}(R)
        -{\delta R}\phi f^{\prime\prime}(R)
        + \cdot\cdot\cdot
        )
        ,
    \end{eqnarray}
the variation of $S_{R}+S_{GHY mod}$ gives
    \begin{eqnarray}\label{f(R) action variation 3}
        \delta S_{R}+\delta S_{GHY mod}
        &&
        =
        \int{d^{4}x}\sqrt{-g}
        (
        {\delta R}f^{\prime}(R)
        \nonumber\\&&
        -\frac{1}{2}f(R)g_{ab}{\delta g}^{ab}
        )
        \nonumber\\&&
        +
        \oint{d^{3}x}\sqrt{|h|}
        (
        {\delta K}f^{\prime}(R)
        \nonumber\\&&
        +
        K f^{\prime\prime}(R){\delta R}
        )
        ,
    \end{eqnarray}

If the action $S_{R}+S_{GHY mod}$ is to be stationary, the last term must vanish (i.e. setting ${\delta R}=0$ on the boundary). However, in (\ref{f(R) action variation 2}) there is no need to set such a condition on the boundary since the second and higher derivatives of $f(R)$ give us a series of field equations for $f(R)$ and its derivatives that can be gathered to give one field equation for $f(\tilde{R})$.
\newline\newline
In conclusion, the Gibbons-Hawking-York term must be an infinite number of boundary integrals in terms of the first and higher derivatives of $f(R)$ that are added to the total action to cancel the surface terms. The reason that there is only one boundary integral for the Einstein-Hilbert action is because the second and higher derivatives of $f(R)$ vanish. $S_{GHY mod}$ require setting ${\delta R}=0$ because we have not considered the rest of the boundary integrals.

\end{document}